\documentclass[preprint]{aastex701}

\usepackage{graphicx}          % 插入图片
\usepackage{subcaption}        % 子图支持
\usepackage{caption} 
\usepackage[normalem]{ulem}
\usepackage{xcolor}
\usepackage{hyperref} % 导入超链接支持

\begin{document}

\title{Preferential Appearance of H$\alpha$ Moreton Waves along Photospheric Magnetic Network Boundaries}

\author[0009-0009-8432-307X]{C. H. Zhai}
\affiliation{School of Astronomy and Space Science, Nanjing University, Nanjing 210023, People's Republic of China}
\affiliation{Key Laboratory for Modern Astronomy and Astrophysics, Nanjing University, Nanjing 210023, People's Republic of China}
\email{changhaozhai@smail.nju.edu.cn}

\author[0000-0002-9908-291X]{Y. W. Ni}
\affiliation{School of Astronomy and Space Science, Nanjing University, Nanjing 210023, People's Republic of China}
\affiliation{Key Laboratory for Modern Astronomy and Astrophysics, Nanjing University, Nanjing 210023, People's Republic of China}
\email{y.w.ni@smail.nju.edu.cn}

\author[0000-0002-4205-5566]{J. H. Guo}
\affiliation{School of Astronomy and Space Science, Nanjing University, Nanjing 210023, People's Republic of China}
\affiliation{Key Laboratory for Modern Astronomy and Astrophysics, Nanjing University, Nanjing 210023, People's Republic of China}
\email{jinhan.guo@nju.edu.cn}

\author[0000-0002-7289-642X]{P. F. Chen}
\affiliation{School of Astronomy and Space Science, Nanjing University, Nanjing 210023, People's Republic of China}
\affiliation{Key Laboratory for Modern Astronomy and Astrophysics, Nanjing University, Nanjing 210023, People's Republic of China}
\affiliation{State Key Laboratory of Lunar and Planetary Sciences, Macau University of Science and Technology, Macau 999078, People's Republic of China}
\email{chenpf@nju.edu.cn}

\correspondingauthor{P. F. Chen}
\email{chenpf@nju.edu.cn}

\begin{abstract}
Moreton waves are rare chromospheric signatures of large-scale coronal disturbances, often associated with big flares and coronal mass ejections (CMEs). Using high-cadence, full-disk H$\alpha$ spectroscopic observations from CHASE, together with the EUV data from SDO/AIA and magnetograms from SDO/HMI, we analyzed a coronal EUV wave and an H$\alpha$ Moreton wave event associated with a filament eruption on 2024 July 29. The Moreton wave fronts are roughly cospatial with the fast-mode coronal EUV wave fronts, which propagate with a speed of $\sim$600 km s$^{-1}$. By comparing the Moreton wave fronts with photospheric features, we found that they preferentially appear along photospheric supergranule boundaries characterized by 1600 \AA\ bright ridges, concentrated magnetic fields, and convective downflows. It is shown that the H$\alpha$ line profiles at the Moreton wave fronts are systematically redshifted. Gaussian fit yields a systematic downward Doppler velocity of 1.73 km s$^{-1}$. Using the bisector method, we further derived height-dependent Doppler velocities in the chromosphere. While there is an expected tendency for the downward velocity to decrease from 4.12 km s$^{-1}$ in the upper chromosphere to 1.60 km s$^{-1}$ in the lower chromosphere, it is intriguing to see an unexpected velocity enhancement in the lower chromosphere. We conjecture that coronal fast-mode MHD waves experience mode-conversion to slow-mode waves, which propagate along magnetic field lines of the magnetic canopy, forming preferential appearance of Moreton waves at magnetic networks, where the convective downflow contributes to the velocity enhancement in the lower chromosphere.
\end{abstract}

\keywords{\uat{Solar coronal waves}{1995}; \uat{Solar magnetic fields}{1503}; \uat{Solar chromosphere}{1479}; \uat{Solar activity}{1475}}

\section{Introduction}

Moreton waves, also known as Moreton-Ramsey waves, are large-scale, transient propagating disturbances observed in the solar chromosphere \citep{1960AJ.....65U.494M,1960PASP...72..357M}. They appear as dark fronts in the red wing of the chromospheric H$\alpha$ line, but as bright fronts in the blue wing and at the H$\alpha$ line center \citep{1961ApJ...133..935A}. The typical propagation speed of Moreton waves ranges from approximately 500 to more than 1500 km s$^{-1}$, and they can travel distances of up to one solar radius across the solar surface \citep{1960PASP...72..357M,1971ASSL...27..156S,2011PASJ...63..685Z}.

Considering that the typical magnetohydrodynamic (MHD) wave speed in the solar chromosphere is only $\sim$100 km s$^{-1}$, the two observational characteristics described above, i.e., the high propagation speed and the long travel distance, make it unlikely for Moreton waves observed in the chromosphere to be fast-mode MHD waves propagating horizontally within the chromosphere. Otherwise, Moreton waves would be strong shock waves with a Mach number of $\sim$5--15, which should be subject to rapid dissipation in the dense chromospheric plasma, preventing them from maintaining high speeds over large distances. Therefore, \citet{1968SoPh....4...30U} and \citet{1973SoPh...28..495U} proposed that the pressure pulses associated with solar flares generate fast-mode MHD waves in the solar corona, where the fast-mode wave speed is in the range of $\sim$500 to 1500 km s$^{-1}$, and that the skirt of the coronal wave front sweeps across the solar chromosphere, hence Moreton waves are formed as the chromospheric plasma is compressed and forced to move downward. This model can self-consistently account for the high propagation speed, long travel distance, and relatively weak amplitude of Moreton waves \citep{2016GMS...216..381C}.

According to the Uchida model, Moreton waves correspond to the sweeping skirt of a fast-mode MHD wave or shock propagating in the solar corona, with the chromospheric H$\alpha$ Moreton waves representing the footprints of these coronal disturbances. 
Observational support for this interpretation was initially provided by the extreme ultraviolet (EUV) observations from the EUV Imaging Telescope (EIT; \citealt{1995SoPh..162..291D}) aboard the Solar and Heliospheric Observatory (SOHO), and later primarily by the Atmospheric Imaging Assembly (AIA; \citealt{2012SoPh..275...17L}) aboard the Solar Dynamics Observatory (SDO; \citealt{2012SoPh..275....3P}).
In the EIT data, occasional ($\sim$7\%) sharp EUV wave fronts were found to be cospatial with H$\alpha$ Moreton wave fronts \citep{2000SoPh..193..161T,2001ApJ...556..421P,2002ApJ...569.1009B,2011PASJ...63..685Z}. 
In the SDO era, two distinct EUV wave components have been identified in individual eruption events, and the faster one, often interpreted as the fast-mode magnetosonic wavefront, was found to be cospatial with the associated Moreton wave front \citep{2002A&A...394..299V,2005SSRv..121..201C,2012ApJ...745L..18A,2012ApJ...752L..23S,2019ApJ...883...32C,2019ApJ...882...90L,2023ApJ...949L...8Z}. Such observational results are consistent with the prediction of \citet{2002ApJ...572L..99C,2005SSRv..121..201C} that there should exist two physically different types of EUV waves, and that the faster component is related to H$\alpha$ Moreton waves.

Despite its success in explaining various properties of H$\alpha$ Moreton waves and their coronal counterparts, the Uchida model was subsequently improved by various authors. First, \citet{cliv99} and \citet{2002ApJ...572L..99C} argued that the coronal counterparts of H$\alpha$ Moreton waves correspond to the piston-driven shock waves excited by coronal mass ejections (CMEs), rather than being generated directly by solar flares. An indicative piece of evidence was revealed by \citet{2006ApJ...641L.153C}, who found that without a CME, even X-class flares are not associated with EUV waves in the corona, not to mention H$\alpha$ Moreton waves. Second, it was argued that inclined eruptions facilitate the appearance of H$\alpha$ Moreton waves even when the piston-driven shock is not very strong \citep{2016SoPh..291...89V}, which was supported by later observations \citep{2019ApJ...882...90L, 2023ApJ...949L...8Z, 2025ApJ...980...42Z}. Moreover, \citet{2016GMS...216..381C} put forward that H$\alpha$ Moreton waves are actually propagating waves in the chromosphere (see their Figure 2, similar to the top panel in our Figure~\ref{fig:fig4}): As a dome-shaped CME-driven shock propagates outward from the source active region, it is refracted into the chromosphere.
Since the fast-mode speed in the chromosphere is approximately an order of magnitude smaller than that in the corona, the wave front in the chromosphere becomes strongly oblique, approaching a near-horizontal orientation according to Snell's law.
Although the wave propagates obliquely through the chromosphere with a speed of $\sim$100 km s$^{-1}$ (typical fast-mode wave speed in the chromosphere), the intersection between the wave front and the H$\alpha$ formation height moves horizontally with a speed the same as the fast-mode wave in the corona, i.e., 500--1500 km s$^{-1}$.
This modification naturally explains why Moreton waves are associated with downward motions of the chromospheric plasma, as the refraction redirects the wave vector nearly downward within the chromosphere.

Based on the slightly modified common understanding of Moreton waves, a CME piston-driven shock wave propagates outward with an inclined-dome-shaped front in the corona. When the wave reaches the solar transition region, it is refracted into the chromosphere, forming a nearly horizontal wave front in the chromosphere. The continued propagation of this oblique wavefront in the chromosphere is expected to produce the H$\alpha$ Moreton wave signatures observed on the solar surface. However, when we analyzed the Moreton wave event on 2024 July 29, we found that the H$\alpha$ Moreton wave fronts do not appear continually across the solar disk, but preferentially appear along the boundaries of the photospheric magnetic networks. Hence it is interesting to explore the relationship between H$\alpha$ Moreton waves and the magnetic structure of the lower solar atmosphere. 
The observational data are described in Section~\ref{sec:2}, the results are presented in Section~\ref{sec:3}, which are discussed in Section~\ref{sec:4}. The main conclusions are summarized in Section~\ref{sec:5}

\section{Observation and Data Analysis} \label{sec:2}
At $\sim$12:47 UT on 2024 July 29, an intermediate filament erupted near the eastern boundary of NOAA active region (AR) 13762, which led to an M8.7-class flare and a fast CME at 13:25 UT. In the northern direction, the Large Angle Spectroscopic Coronagraph \citep[LASCO,][]{brue95} aboard SOHO revealed that the CME exhibited the typical three components, i.e., a leading frontal loop, a cavity, and a bright core. The projected speed of the frontal loop was 626 km s$^{-1}$. In addition, a halo front was observed around the CME with a projected speed of $\sim$1000 km s$^{-1}$, which is likely the CME-driven shock wave \citep{Chen2011}. As the filament erupted, a coronal wave was observed to propagate northward in the EUV images. Notably, this coronal EUV wave was accompanied by a chromospheric H$\alpha$ Moreton wave.

The filament eruption, the associated solar flare, and the coronal EUV wave were observed by the AIA telescope aboard the SDO satellite. The AIA telescope has seven EUV channels, two UV channels, and one white-light channel, each of which provides full-disk images of the Sun with a pixel size of $0\farcs6$ and a cadence of 12 s.
To examine the spatial relationship between Moreton wave fronts and the photospheric network structures, we used the data from AIA 1600 \AA\ images and from the Helioseismic and Magnetic Imager (HMI; \citealt{2012SoPh..275..207S}), which provides photospheric magnetograms and Dopplergrams with a pixel size of $0\farcs5$ and a cadence of 45 s. We processed the HMI Dopplergrams using the \texttt{hmi\_clean} software package \citep{2021ApJ...916...87K}. Systematic removal of large-scale artifacts and instrumental effects allowed us to isolate the convective flows and other fine-scale features on the solar surface. An average of 20 min Dopplergrams was taken in order to derive the convective flow, smearing out $p$-mode oscillations.

To investigate the Moreton waves, we analyzed the H$\alpha$ spectral data from the Chinese H$\alpha$ Solar Explorer (CHASE) mission, which was launched on 2021 October 21. 
The CHASE data provide two-dimensional spectroscopic observations of the full solar disk with a pixel size of $1\farcs04$ in the binning mode and a cadence of 73 s \citep{Li2022, 2022SCPMA..6589603Q}. The spectra consist of the H$\alpha$ line sampled at 118 wavelength positions over a spectral range from 6559.7 to 6565.9~\AA, as well as the \ion{Fe}{1} line over a spectral range from 6567.8 to 6570.6~\AA.

\begin{figure}[htbp]
    \centering
    \includegraphics[width=0.8\textwidth]{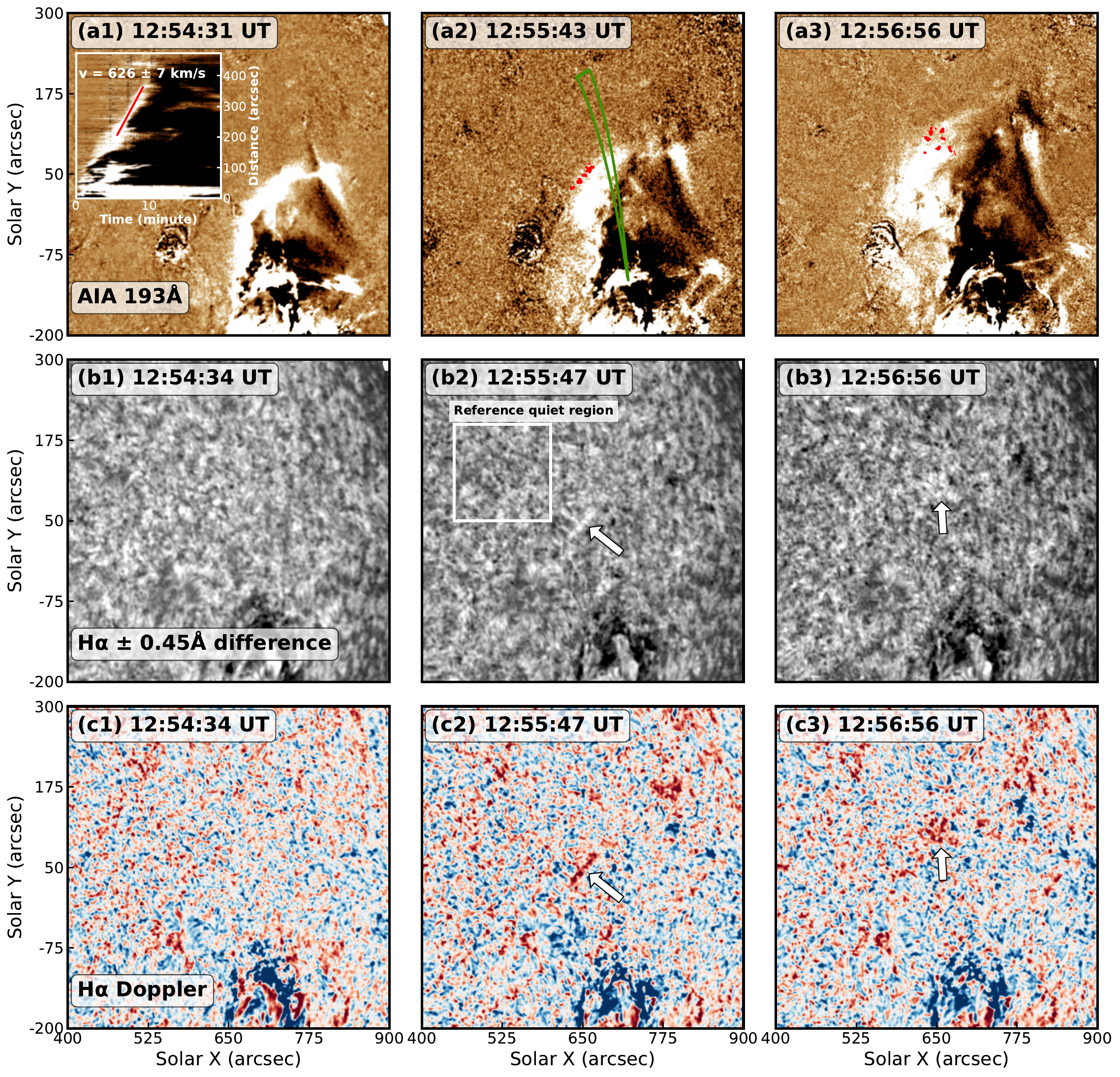} % 图片路径
    \caption{Evolution of the coronal EUV wave and the associated H$\alpha$ Moreton wave.
Panels (a1)--(a3) show time sequences of the base-difference images in AIA 193~\AA, illustrating the propagation of the coronal EUV wave. The base image is at 12:51:03 UT. The time-distance diagram in panel (a1) is constructed along the sector marked by the green line in panel (a2). The red markers in panels (a2--a3) indicate the locations of the contemporaneous Moreton wave fronts. Panels (b1)--(b3) display the difference H$\alpha$ images constructed from the blue-wing minus red-wing intensity, highlighting the chromospheric signatures of the Moreton wave as indicated by the arrows. Panels (c1)--(c3) show the corresponding H$\alpha$ Dopplergrams, where the redshifted features indicated by the arrows mark the propagation of the Moreton wave fronts. The animation of this figure is available online. The animation of this figure is available online.}

    \label{fig:fig1} % 图片标签（用于引用）
\end{figure}

\section{Results} \label{sec:3}
As displayed in Figures \ref{fig:fig1}(a1)--(a3), the solar flare associated with a filament eruption occurred near the northern edge of the source active region at the helioprojective coordinates of (721$\arcsec$, -183$\arcsec$) on 2024 July 29. 
Around the impulsive phase of the solar flare, at 12:50:41 UT, a coronal EUV wave emanated from the source active region on the northern side. 
At 12:55 UT when the flare reached its peak, the EUV wave became significantly brighter.
Figures~\ref{fig:fig1}(a1)--(a3) depict three snapshots of the EUV 193~\AA\ base-difference images, where the EUV wave is seen to propagate northward. 
This preferential northward propagation is due to the fact that the active region with much stronger magnetic field is to the south of the eruption site, consistent with the results of \citet{zhong2025magnetic}. To estimate the propagation speed of the EUV wave, we selected a slice extending from the source region along the direction of wave propagation (indicated by the green sector in Figure~\ref{fig:fig1}(a2)). The corresponding time-distance diagram is shown in the inset of Figure~\ref{fig:fig1}(a1). The propagation speed was derived by tracking the leading edge of the wave front in the diagram, yielding a value of $626\pm7$ km s$^{-1}$. Since the bright ridge in the time-distance diagram is almost straight, there is no evident acceleration or deceleration during the interval, and the error bar in the fitting is small.
%A detailed time-distance analysis indicates that the propagation speed of the EUV wave was 626$\pm$7 km s$^{-1}$.

At the same time, the CHASE satellite detected an H$\alpha$ Moreton wave. The middle row of Figure~\ref{fig:fig1} displays the difference filtergram obtained by subtracting the  intensity at H$\alpha$+0.45 \AA\ from that at H$\alpha$-0.45 \AA. The Moreton wave fronts are indicated by the white arrows. Although the EUV wave is clearly visible at 12:54:37 UT (Figure~\ref{fig:fig1}(a1)), the corresponding Moreton wave front is not yet detectable at this time (Figure~\ref{fig:fig1}(b1)). As the disturbance propagated northward, the Moreton wave front became prominently visible at 12:55:47 UT (Figure~\ref{fig:fig1}(b2)). Approximately one minute later, at 12:56:56 UT, the Moreton wave front intensified further, appearing more prominent and wider in space. 
Dividing the travel distance by the time difference, the propagation speed of the Moreton wave is estimated to be 599$\pm$5 km s$^{-1}$.
To compare the spatial relationship between the Moreton wave and the coronal EUV wave, we overlaid the Moreton wave fragments onto the EUV wave images in Figures~\ref{fig:fig1}(a2)--(a3) marked by the red contours. It is evident that the Moreton wave is almost cospatial with the leading edge of the coronal EUV wave. Note that the time difference between the H$\alpha$ images and the EUV images is 4 s for the second column, i.e., the simultaneous Moreton wave front should be shifted southward by $4\times 599=2,396$ km in Figure \ref{fig:fig1}(a2), corresponding to $3\farcs25$. Following the method mentioned in \citet{2019ApJ...882...90L}, we estimated the Mach number of the EUV wave, which is 1.15. The appearance of the Moreton wave is attributed to the inclined filament eruption.
%Note that the time difference between the H$\alpha$ images and the EUV images is 17 s for the second column and 14 s for the third column, i.e., the simultaneous Moreton wave front should be shifted southward by $17\times 599=10.2\times 10^3$ km and $14\times 599=8.4\times 10^3$ km, corresponding to 13.8\arcsec and 11.4\arcsec, respectively.

The bottom row of Figure~\ref{fig:fig1} shows the evolution of Doppler velocity of the chromosphere, where the Doppler velocity was derived via the classical cross-correlation technique using the whole H$\alpha$ line profile \citep[see][]{2022SCPMA..6589603Q}. As indicated by the white arrows, the Moreton wave fronts were characterized by redshifts. Similar to the evolution of the H$\alpha$ wing intensity, the Doppler redshifts were very significant when the Moreton wave fronts were clearly visible at 12:55:47 and 12:56:56 UT. It is noted that the flare source region is predominantly characterized by blueshifts, whilst small-scale fragmented redshift patches are still present (Figures~\ref{fig:fig1}(c1)--(c3)). Despite the cospatiality between the Moreton wave and EUV wave, it is noted that while the coronal EUV wave expanded radially, the trajectories of the Moreton wave fronts did not follow a great circle on the solar surface. This discrepancy can be attributed to the much more inhomogeneous and highly structured magnetic fields in the lower atmosphere compared to those in the corona.

\begin{figure}[htbp]
    \centering
    \includegraphics[width=1\textwidth]{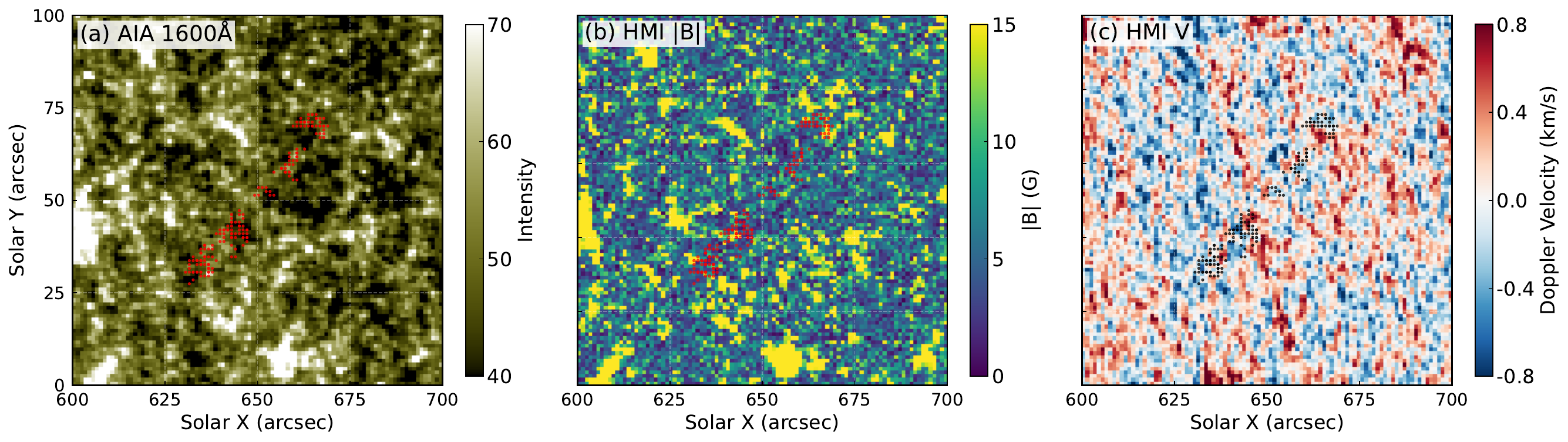} % 图片路径
    \caption{
Spatial relationship between the Moreton wave front at 12:55:47 UT and different proxies of photospheric networks, i.e., the AIA 1600~\AA\ intensity image (panel a), the HMI longitudinal magnetic intensity (panel b), and the HMI Dopplergram. The Moreton wave front is indicated by the red dots in panels (a) and (b), and black dots in panel (c).}      
    \label{fig:fig2} % 图片标签（用于引用）
\end{figure}

Hence, we compared the spatial relationship between the Moreton wave and photospheric networks, i.e., supergranules, which are characterized by enhanced emission in AIA 1600~\AA\ images, concentrated magnetic elements, and convection downflows. The three panels of Figure~\ref{fig:fig2} display the AIA 1600~\AA\ intensity map (panel a),  the HMI longitudinal magnetogram (panel b), and the HMI Dopplergram (panel c) at 12:55:47 UT, where the Moreton wave front at the same time is overlaid as the red dots in panels (a)--(b) and black dots in panel (c). 
It is noteworthy that the Moreton wave front is cospatial with the photospheric network boundaries, which are characterized by bright ridges in AIA 1600~\AA\ images, magnetic elements, and photospheric downflows, as indicated by the redshifts in the HMI Dopplergram.

\begin{figure}[htbp]
    \centering
    \includegraphics[width=0.9\textwidth]{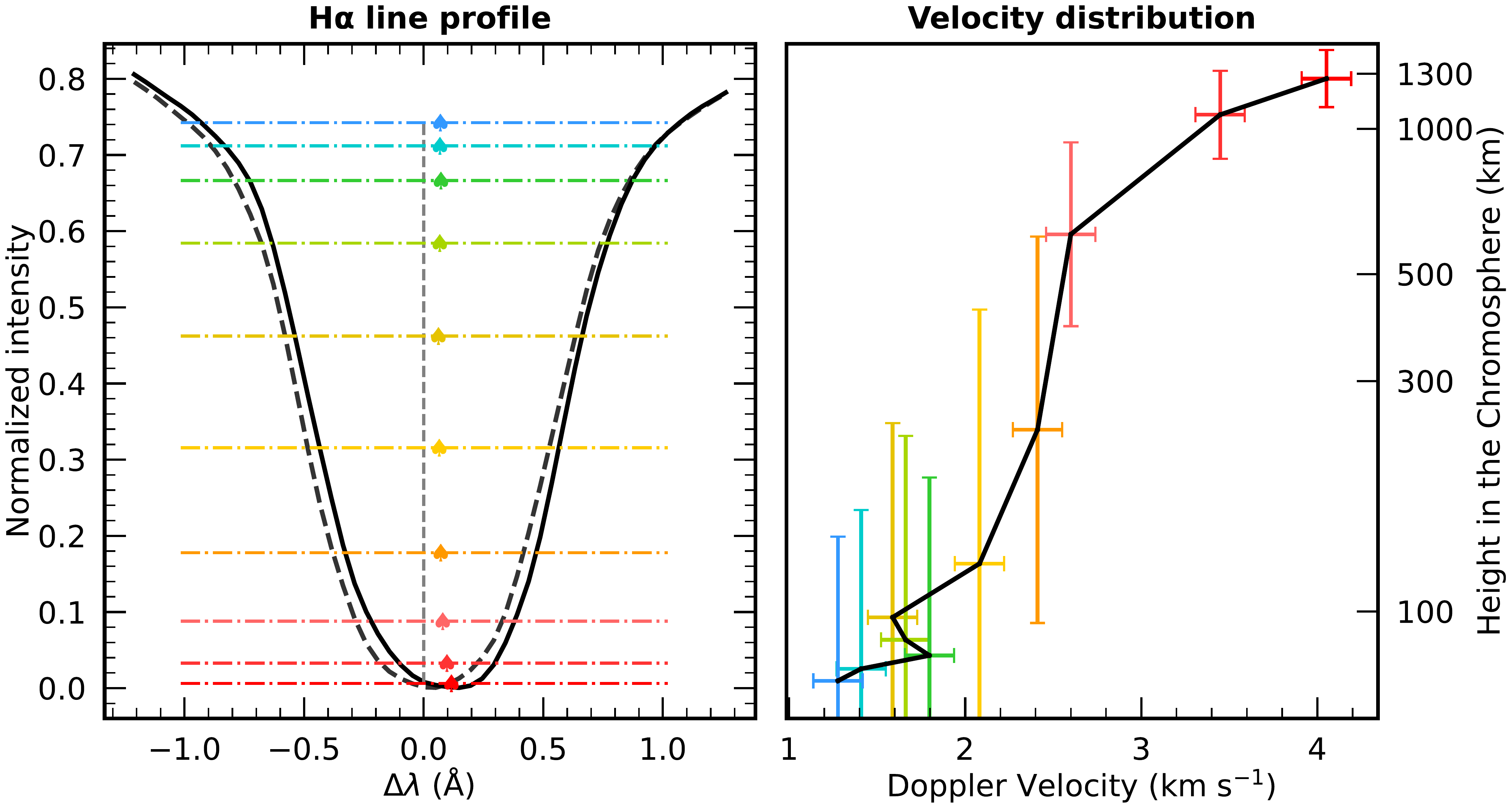}
    \caption{Left panel: Normalized H$\alpha$ spectral profile of the Moreton wave front at 12:55:47 UT (solid black line) and the spectral profile of the reference quiet region (dashed black line), where the vertical dashed gray line indicates the line center in the rest frame of reference. The colored horizontal lines (from bottom to top) mark the wavelength windows from $\pm$0.1~\AA\ to $\pm$1.0~\AA. The colored dots indicate the bisector offset of each wavelength window (corresponding to certain heights in the chromosphere).  Right panel: Derived Doppler velocity of the Moreton wave as a function of chromospheric height based on the bisector method.}
    \label{fig:fig3} % 图片标签（用于引用）
\end{figure}

The CHASE mission provides high-resolution H$\alpha$ spectra across the entire solar disk. The H$\alpha$ line profile of the Moreton wave front is shown as the solid black curve in the left panel of Figure~\ref{fig:fig3}, representing the spatially averaged spectrum over the Moreton wave front region identified from Figure~\ref{fig:fig1}(b2) at 12:55:47~UT. Note that this spectral profile is an average of 138 pixels, which satisfy two criteria: (1) the difference intensity between H$\alpha$-0.45\AA\ and H$\alpha$+0.45\AA\ exceeds the background level by more than three standard deviations; (2) the whole Doppler velocity is larger than 1 km s$^{-1}$.
For comparison, we overplot the averaged H$\alpha$ line profile (dashed black line) from a nearby quiet region indicated by the white rectangle in Figure~\ref{fig:fig1}(b2), including 39,600 pixels. To ensure that the signal is sufficiently strong while avoiding overshooting effects from convection motions, this region is selected specifically. It is seen from the left panel of Figure~\ref{fig:fig3} that the H$\alpha$ line profile associated with the Moreton wave is red shifted, as expected with the current understanding of Moreton waves.
The H$\alpha$ spectral line, from the line center to the line wings, is formed from the upper chromosphere to the lower chromosphere. As an approximation, we can derive the Doppler velocity at different heights in the lower solar atmosphere by applying the traditional bisector method to the H$\alpha$ spectral profile \citep{1981A&A....96..345D,1987A&A...173..161C}. The bisector offsets at different wavelength windows of the H$\alpha$ line at the Moreton wave front are marked as the colored dots in the left panel of Figure~\ref{fig:fig3}, and the dashed vertical gray line denotes the H$\alpha$ line center in the reference region. It is straightforward to see that the H$\alpha$ spectrum at the Moreton wave front is differentially redshifted at different chromospheric altitudes. 
Using the approximate formation heights of the H$\alpha$ line profile estimated with the method described in \citet{2012ApJ...749..136L} and \citet{2024NatAs...8.1102R}, we derived the altitude-dependent distribution of the Doppler velocities of the chromosphere, which is plotted in the right panel of Figure~\ref{fig:fig3}. It is revealed that in the upper chromosphere, the downward velocity is the maximum, reaching up to 4.12 km s$^{-1}$. As expected, the amplitude of the bisector velocity decreases from the upper chromosphere toward the lower chromosphere since the plasma density increases drastically. However, this trend reverses below an altitude of approximately 97 km, where the downward velocity begins to increase again from the lower chromosphere toward the solar surface.

%With the approximate formation height of the H$\alpha$ line \citep{2012ApJ...749..136L,2024NatAs...8.1102R}, we can derive the distribution of the Doppler velocity of the chromosphere along with the altitude, which is plotted in Figure 3(b).
%It is seen that at the upper chromosphere, the downward velocity reaches up to~4 km s$^{-1}$. As expected, the bisector velocity decreases from the upper chromosphere toward the lower chromosphere.
%Hence, it is also noticeable that such a tendency beings to reoccur below the altitude of ?? km. That is to say, the downward velocity starts to increase from the lower chromoshpere toward the solar surface.
%With the approximate formation height of the H$\alpha$ line, we derive the altitude-dependent distribution of Doppler velocities in the chromosphere, as plotted in Figure 3(b). It is evident that in the upper chromosphere, the downward velocity reaches 4km s$^{-1}$. As expected, the magnitude of the downward velocity decreases from the upper chromosphere to the lower chromosphere.
%Furthermore, it is noticeable that this trend reverses below an altitude of approximately xx km. That is to say, the downward velocity begins to increase again from the lower chromosphere toward the solar surface.

\section{Discussion} \label{sec:4}

\subsection{Height-distribution of Doppler velocity in the Moreton wave}

It is well known that Moreton waves are formed due to downward motion of the solar chromosphere compressed by coronal shock waves \citep{1968SoPh....4...30U,2016GMS...216..381C}. As a result, Moreton waves manifest as bright fronts at H$\alpha$ line center and blue wing or dark fronts at H$\alpha$ red wing. For the first time, we present the high spectral resolution observation of a Moreton wave obtained by the CHASE mission. As shown in the left panel of Figure~\ref{fig:fig3}, the H$\alpha$ line profile at the Moreton wave front is clearly redshifted toward longer wavelength. 
A Gaussian fit to the whole line profile yields a redshift of 0.038~\AA, corresponding to a Doppler redshift of 1.73 km s$^{-1}$. These values are comparable to those obtained by \citet{2007ApJ...658.1372B} using an independent Doppler measurement, indicating the consistency between the CHASE measurements and previous methods. However, it is noticed that the H$\alpha$ line is not shifted as a whole, and different wavelengths are shifted differentially. Owing to the high spectral resolution of CHASE, we are able to derive the height-dependent Doppler velocity associated with the Moreton wave based on the bisector method. As shown in the right panel of Figure~\ref{fig:fig3}, the downward speed reaches $\sim$4.12$\pm$0.14 km s$^{-1}$ in the upper chromosphere. The Doppler velocity decreases toward lower altitudes in response to the increasing atmospheric density. 
It is seen that at a height of 97 km above the photosphere, the downward velocity is reduced to only 1.60$\pm$0.14 km s$^{-1}$. Surprisingly, the Doppler velocity begins to slightly increase from 1.60$\pm$0.14 km s$^{-1}$ at a height of 97 km to 1.79$\pm$0.14 km at a height of 81 km.
Below this height range, the downward velocity decreases drastically toward the photosphere. We examined various locations along the Moreton wave front and found that all sampled pixels exhibit a similar trend: the downward velocity decreases from the upper chromosphere toward the lower chromosphere. However, below an altitude of approximately 97 km above the photosphere, the downward velocity shows a slight increase with decreasing height before decreasing again at lower altitudes. 

It is generally believed that Moreton waves correspond to the downward motion of the chromosphere driven by coronal shock waves. From this physical perspective, the downward velocity is expected to decrease with decreasing altitude as the plasma density increases. Therefore, the weak velocity enhancement observed at lower chromoshpere is unlikely to be directly related to the Moreton wave itself. 
%at the heights of approximately 404 to 289 km is unlikely to be directly related to the Moreton wave itself. 
We speculate that this feature may instead be associated with localized downdrafts occurring at the boundaries of the photospheric network \citep{1989ApJ...343..475T, 2000SoPh..193..299H}, and a little overshooting leads to the enhanced downward velocity in the lower chromosphere. 
To confirm this hypothesis, we compared the spatial relationship between the Moreton wavefronts and the Dopplergram of the photosphere obtained from HMI at 12:55:47 UT. As shown in Figure~\ref{fig:fig2}(c), the Moreton wave front coincides spatially with the redshifts at the boundaries of photospheric networks.
%To confirm this hypothesis, we compared the spatial relationship between the Moreton wavefronts and the magnetic intensity (absolute values) of the HMI line-of-sight magnetogram at 12:55:47 UT. As shown in Figure \ref{fig:fig2}, the Moreton wave front is indeed spatially coincident with the AIA 1600 \AA\ brightenings and the concentrated magnetic elements at the boundaries of photospheric networks.

\subsection{Implication for the formation mechanism of Moreton wave}

\begin{figure}[htbp]
    \centering
 %       \begin{subfigure}[b]{0.9\textwidth}
 %       \centering
 %       \includegraphics[width=\linewidth]{fig41.PNG}
 %       \label{fig:sub1}
 %   \end{subfigure}
 %   
 %   \vspace{0.3cm}  % 调整上下子图的间距
 %   
 %   \begin{subfigure}[b]{0.8\textwidth}
 %       \centering
 %       \includegraphics[width=\linewidth]{fig42}
 %       \label{fig:sub2}
 %   \end{subfigure}
    \includegraphics[width=12cm]{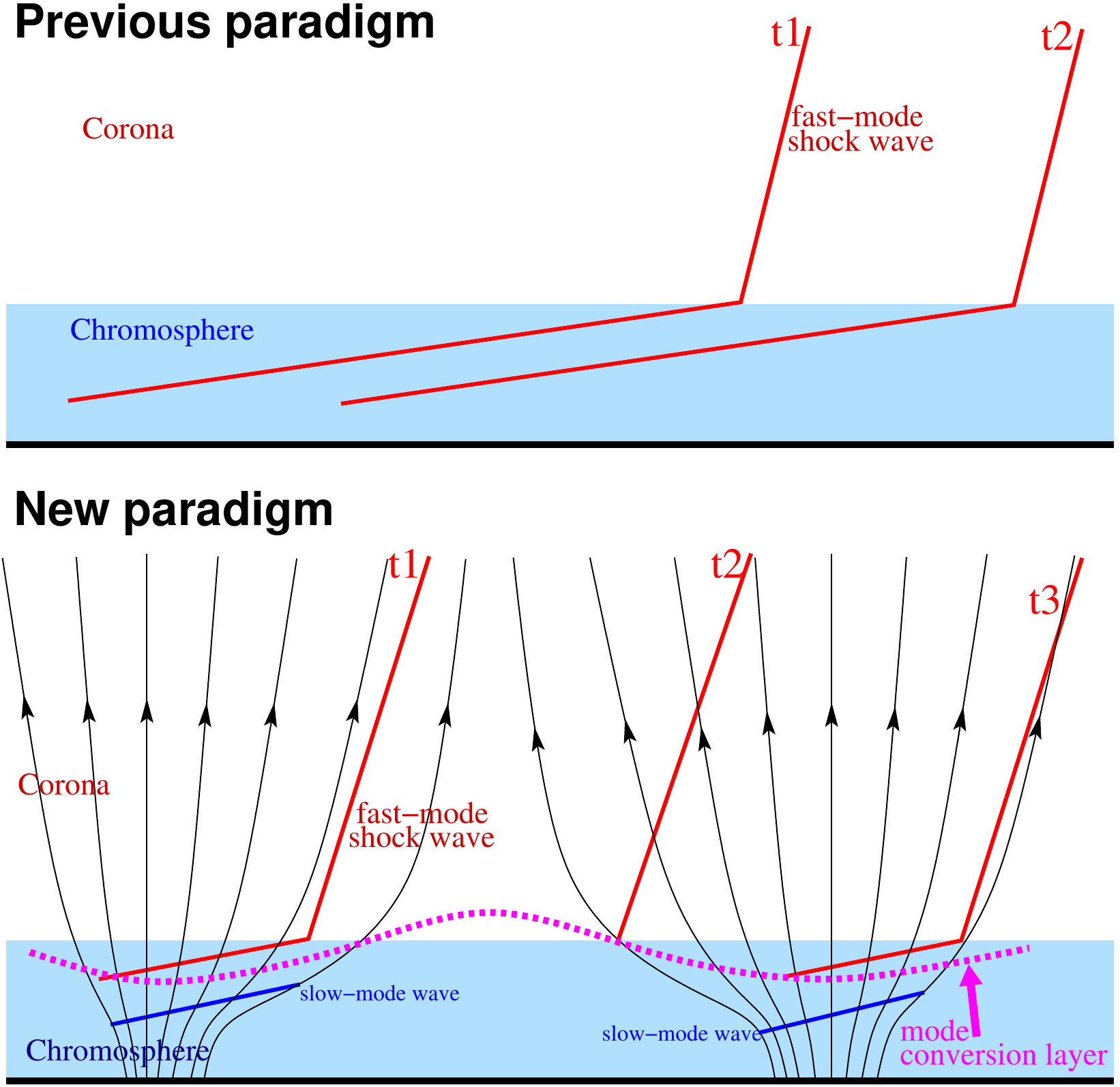}
    \caption{Schematic sketch comparing the traditional understanding of Moreton wave and our new paradigm. The top panel is the old paradigm presented in \citet{2016GMS...216..381C}, the bottom panel is the new paradigm considering the strong inhomogeneity of the magnetic field in the lower solar atmosphere.
    The thin black lines represent the magnetic canopy-like structure. The red lines indicate the fast-mode MHD wave front whereas the blue lines correspond to slow-mode wave fronts in the chromosphere.}
    \label{fig:fig4}
\end{figure}

Moreton waves were discovered before CMEs, hence it has been taken for granted that Moreton waves are generated by the pressure pulse in solar flares, and Moreton waves were sometimes called flare waves \citep{1968SoPh....4...30U, warm04}. 
According to the Uchida model, the pressure pulse in the flare loops generates fast-mode shock waves in the corona, whose footpoints sweep the chromosphere so as to generate Moreton waves. Although it was later argued that the coronal shock waves should be CME piston-driven, rather than the flare pressure pulse-driven \citep{cliv99, 2002ApJ...572L..99C, 2007ApJ...658.1372B}, the essence of the Uchida model, i.e., the coronal shock wave associated with solar eruptions sweeps and compresses the solar chromosphere, still holds true.
However, such a traditional model does not explicitly address the detailed physical process how the chromosphere is compressed by the coronal shock wave. 

\citet{2004ApJ...610..572G} proposed that the perturbations in the downstream of the coronal shock wave generate slow-mode waves, which propagate down along magnetic field lines from the corona to the chromosphere. 
On the other hand, \citet{2016GMS...216..381C} argued that the fast-mode shock wave is refracted from the corona to the chromosphere. Since the fast-mode wave speed is about one order of magnitude smaller in the chromosphere than in the corona, the wave front in the chromosphere intersects with the solar surface with a small angle, as illustrated by the red lines in the top panel of Figure~\ref{fig:fig4}. In this paradigm, the strong inhomogeneity of the magnetic field in the lower solar atmosphere was not considered, therefore it was taken for granted that the magnetic field changes smoothly and the Moreton wave front propagates outward in the chromosphere in a continual way. However, in this Letter, we found that Moreton waves prefer to appear above the boundary of supergranulations, rather than propagating continually.

To explain such a phenomenon, we noted that the magnetic field in the lower solar atmosphere is highly structured. Different from the corona, which is permeated with smoothly distributed magnetic field, the magnetic flux tubes in the photosphere are swept and concentrated toward the supergranule boundaries. As a result, the magnetic configuration in the lower solar atmosphere forms a canopy-like structure \citep{1976RSPTA.281..339G}, where magnetic field is concentrated at network boundaries, leaving a nonmagnetic canopy immediately above the supergranule cells. Within such a configuration, a horizontally propagating fast-mode coronal wave may not be able to generate chromospheric imprints continually, but leave chromospheric imprints on the network boundaries. The mechanism is elaborated as follows: The plasma $\beta$, i.e., the ratio of thermal to magnetic pressure, changes from much less than unity in the low corona to much larger than unity in the photosphere. Consequently, there exists an equipartition layer where the Alfv\'en speed equals sound speed, i.e., $\beta=2/\gamma$ (where $\gamma=5/3$), a condition under which mode conversion between fast- and slow-mode MHD waves is expected to be efficient \citep{call05, 2016GMS...216..381C, chan18}. In such a layer, an incident fast-mode wave can partially convert into a slow-mode wave, which is compressive in nature and propagate preferentially along magnetic field lines into the chromosphere. For the canopy-like magnetic configuration shown in the lower panel of Figure~\ref{fig:fig4}, the $\beta=2/\gamma$ layer resides in the chromosphere for the strongly magnetized patches and in the low corona above the nonmagnetized areas, as indicated by the dashed magenta line in the lower panel of Figure \ref{fig:fig4}.

As shown in the lower panel of Figure \ref{fig:fig4}, for a CME piston-driven shock wave (red thick lines) propagating horizontally, if the mode conversion layer is inside the chromosphere, the fast-mode shock wave would first be refracted into the chromosphere, forming a nearly horizontal fast-mode shock wave. When the fast-mode wave crosses the equipartition layer, most of the fast-mode wave would be converted to a slow-mode wave (blue lines), as simulated by \citet{2003ApJ...599..626B}. If the mode conversion layer is in the low corona, the fast-mode coronal wave would be similarly converted to a slow-mode wave before being refracted to the chromosphere. In either case, the resulting slow-mode waves (blue lines) can propagate only along magnetic field, they would follow the converging magnetic field lines, and be concentrated at the magnetic network boundaries. The resulting slow-mode waves generate field-aligned velocity perturbations in the chromosphere, leading to detectable H$\alpha$ redshift signatures, whereas not much perturbation leaks to the interiors of the networks. It is noted that the magnetic network simply channels the fast-mode piston-driven shock wave from the corona, and it does not guarantee the appearance of Moreton waves in the chromosphere. Along the coronal shock front, only the strongest portion would generate Moreton waves \citep{2016SoPh..291...89V}, which is why the Moreton wave front at any instance is much narrower than the coronal EUV wave front.

If the above interpretations are correct, the Moreton wave fronts would be narrower in the H$\alpha$ line wing filtergrams than at the H$\alpha$ line center filtergrams, since the line wings are formed in the lower chromosphere where magnetic flux tubes are more contracted than in the upper chromosphere. It is interesting to see that the CHASE data are consistent with such a deduction. We therefore suggest that the preferential appearance of Moreton waves along photospheric magnetic network boundaries reflects the critical role of local magnetic topology and plasma conditions in mediating the chromospheric response to the initial coronal fast-mode waves. 
Our results refined the classical interpretation of \citet{1968SoPh....4...30U} by emphasizing that the chromospheric response to a coronal fast-mode wave is highly inhomogeneous due to the structured magnetic environment of the lower solar atmosphere.

\section{Summary} \label{sec:5}

In this work, we presented a detailed observational study of an H$\alpha$ Moreton wave associated with the 2024 July 29 solar eruption event, using the full-disk spectroscopic observations from the CHASE mission in combination with SDO/AIA and SDO/HMI data. The Moreton wave, along with the fast-mode coronal EUV wave, propagates on the solar surface with a speed of $\sim$600 km s$^{-1}$. This event provides an opportunity to investigate the fine structure and formation mechanism of Moreton waves with unprecedented spectral diagnostics. Our main results are summarized as follows.

First, we found in this event that the Moreton wave does not propagate outward continually. Instead, it preferentially appears at photospheric network boundaries, which are characterized by concentrated magnetic elements and persistent downflows. We conjecture that there exists wave mode conversion at the equipartition layer where Alfv\'en speed equals sound speed. Because of the magnetic canopy structure, the equipartition layer is located in the chromosphere at the photospheric networks and in the low corona inside the networks. When a coronal fast-mode wave (or shock wave) crosses the equipartition layer, most of the fast-mode wave is converted to slow-mode wave, which then propagates along the magnetic field lines. As a result, the perturbation becomes concentrated at photospheric networks, i.e., supergranules.

Second, the Moreton wave fronts are associated with height-dependent Doppler redshifts. We derived the height-dependent Doppler velocity distribution of the chromosphere by applying the bisector method to the H$\alpha$ line profile obtained with CHASE. It is expected to see that the downward velocity, which is $\sim$4.12 km s$^{-1}$ in the upper chromosphere, decreases toward lower altitudes. However, it is intriguing to find that the downward velocity begins to increase from the height of 
97 km to 81 km above the photosphere before decreasing again lower down. We attribute this local enhancement to the convective downdraft of supergranules.

Our findings highlight the importance of considering canopy-like magnetic structure in interpreting the observational features of Moreton waves, which should be further explored with MHD simulations, together with high-resolution spectroscopic observations.

\begin{acknowledgments}
The authors thank Shihao Rao for his assistance in processing the CHASE data.
This research was supported by NSFC (12127901 and 1250030413), the National Key Technologies Research and Development Program of the Ministry of Science and Technology of China (2020YFC2201200), and the China National Postdoctoral Program for Innovative Talents fellowship (BX20240159).
\end{acknowledgments}

\bibliography{ref}{}
\bibliographystyle{aasjournalv7}



\end{document}